\documentclass[12pt,a4paper]{article}

\usepackage{amsmath}

\def\up{\ensuremath{\vert\hspace{-0.5ex}\uparrow\rangle}}
\def\down{\ensuremath{\vert\hspace{-0.7ex}\downarrow\rangle}}
\def\upC{\ensuremath{\langle\uparrow\hspace{-0.7ex}\vert}}
\def\downC{\ensuremath{\langle\downarrow\hspace{-0.65ex}\vert}}

\def\p{\ensuremath{\mathcal{P}}}
\def\t{\ensuremath{\mathcal{T}}}
\def\pt{\ensuremath{\mathcal{PT}}}
\def\c{\ensuremath{\mathcal{C}}}
\def\hpt{\ensuremath{H_{\mathcal{PT}}}}
\def\ptqm{\ensuremath{\pt}-QM}

\begin{document}
\title{Is \pt{}-symmetric quantum mechanics just quantum mechanics in a non-orthogonal basis?}
\author{Damien Martin\footnote{email: djmartin@ucdavis.edu}\\\emph{\scriptsize Department of Physics, University of California at Davis, 1 Shields Avenue, Davis, CA 95616}}
\date{}

\maketitle


\begin{abstract}
One of the postulates of quantum mechanics is that the Hamiltonian is Hermitian, as this guarantees that the eigenvalues are real. Recently there has been an interest in asking if $H^\dagger = H$ is a necessary condition, and has lead to the development of \pt-symmetric quantum mechanics. This note shows that any finite physically acceptable non-Hermitian Hamiltonian is equivalent to doing ordinary quantum mechanics in a non-orthogonal basis. In particular, this means that there is no experimental distinction between \pt{}-symmetric quantum mechanics and ordinary quantum mechanics for finite systems. In particular, the claim that \pt-symmetric quantum mechanics allows for faster evolution than Hermitian quantum mechanics is shown to be a problem of physical interpretation.
\end{abstract}

\section*{Introduction}
One of the postulates of quantum mechanics is that the Hamiltonian is Hermitian. This is a sufficient condition for real eigenvalues and unitary time evolution. Recently the question of whether or not this condition is necessary has been raised, and a class of Hamiltonians called \emph{unbroken \pt-symmetric} Hamiltonians are shown to have real eigenvalues, where \p{} and \t{} are the parity and time-reversal operators respectively \cite{introPT, discussPT}. To make time evolution unitary the inner product is determined by the Hamiltonian. Because of its novelty and some confusion in the foundations of the field, the first section of this paper will outline the major results of \pt{}-symmetric quantum mechanics (\ptqm).

\paragraph{}
This observation raises two serious questions. The first is if this actually generalises quantum mechanics at all, or if each member of this new class of Hamiltonians is somehow equivalent to a Hermitian Hamiltonian. The second question is if these new Hamiltonians are different, why have we not encountered a system which requires a \pt{}-symmetric Hamiltonian? Naturally, without knowing something about the possible spectrums of all \pt{}-symmetric Hamiltonians and all Hermitian Hamiltonians the second question is difficult to answer, as it may be the case that any \pt{}-symmetric Hamiltonian can be approximated arbitrarily well by a Hermitian Hamiltonian. In such a case, the fact that we are not forced to look at \pt{}-symmetry experimentally would be no more surprising than the fact we do not experimentally find irrational numbers. 

\paragraph{}
This paper answers these questions, at least for finite dimensional systems. For systems with a \emph{finite}-dimensional Hilbert space it is shown ordinary quantum mechanics and \ptqm{} are equivalent for a physically reasonable theory. The criteria for physical reasonableness are defined in section \ref{sec:PT}. It is shown that a non-Hermitian Hamiltonian corresponds to a quantum system written in a non-orthogonal basis. By a basis transformation, the two systems are equivalent and agree for all observables. The second question does not even arise as no system \emph{requires} a Hermitian Hamiltonian, and conversely any system we already know about could be reformulated in a non-orthogonal basis (and hence give rise to non-Hermitian quantum mechanics).

\paragraph{}
This result would preclude an experimental distinction between finite-dimensional ordinary quantum mechanics and \ptqm, as the two are equivalent. Recently it was claimed that one could find a \pt-symmetric Hamiltonian that would evolve between two fixed states in an arbitrarily short amount of time,  subject to the constraint that the difference between the greatest and least eigenvalues $E_+ - E_-$ is held fixed \cite{FastBender}. In contrast, there is a non-trivial fastest time evolution between states of a Hermitian Hamiltonian \cite{quantumshort}. It was then proposed that this test (referred to as the quantum brachistochrone) would allow one to experimentally distinguish between \ptqm{} and ordinary quantum mechanics. This paper shows that this is not an experimental test, but rather a coordinate artefact. 

\paragraph{}
The organisation of this paper is as follows. In section \ref{sec:PT} \pt-symmetry is introduced and the criteria for a physically reasonable theory are laid out. In section \ref{sec:QMNO} it  is shown how to do quantum mechanics in a non-orthogonal basis, and how this transforms a Hermitian Hamiltonian into a non-Hermitian Hamiltonian. This may seem like a nonsensical statement as the hermiticity of an operator can be defined in a basis independent way, and in appendix A this semantic confusion is addressed. This construction works for any quantum system: finite, countably infinite or even a field theory. It is also shown how the quantum brachistochrone results mentioned in the last paragraph does not qualify as an experimental result. In section \ref{sec:PT2QM} it is shown that \emph{any finite} quantum system is equivalent to an ordinary quantum system governed by a Hermitian Hamiltonian. Possible uses of \pt{}-symmetry are also discussed. Finally in section \ref{sec:further} open research questions are posed.

\section{Primer on \pt{}-symmetry}
\label{sec:PT}

\subsection{Anti-linear operators}
One of the primary motivations for being interested in \ptqm{} is that a more general class of Hamiltonians have real eigenvalues. It is well-known that \p{} is represented by a linear operator but \t{} is represented by an anti-linear operator, thus making the composite operator \pt{} anti-linear. Because $\p^2 = \t^2 = 1$, it is also clear that the possible eigenvalues of \pt{} are $\pm 1$ and can thus never vanish. A system which is symmetric under both parity and time-reversal would be governed by a Hamiltonian \hpt{}, which satisfies the following rule:
\begin{equation}
[\hpt, \pt] = 0.
\end{equation}
In exploring the consequences of this equation, we can generalise slightly to replace \pt{} with an arbitrary anti-linear operator $\mathcal{A}$ that does not possess zero eigenvalues. The operator $\mathcal{A}$ will represent some transformation, and we are interested in the analogous case where $[H_\mathcal{A}, \mathcal{A}] = 0$. We are assuming that $H_\mathcal{A}$ is a linear operator, but we are not assuming that it is Hermitian. 

Take a state $\vert E,a \rangle$ that is an eigenvector of  $H_{\mathcal{A}}$  and $\mathcal{A}$, with eigenvalues $E$ and $a$ respectively. Following the simple proof given in \cite{introPT}, we have
\begin{align}
 H_{\mathcal{A}} \mathcal{A} \vert E, a \rangle &= Ea\vert E,a\rangle\\
 \mathcal{A}  H_{\mathcal{A}} \vert E, a\rangle &= E^*a \vert E,a \rangle.
 \end{align}
 Because  $H_{\mathcal{A}}$  and $\mathcal{A}$ commute these expressions are equal. This can only occur if $Ea=E^* a$, and as by hypothesis $\mathcal{A}$ has no zero eigenvalues we must have $E = E^*$. 

\paragraph{}
However, unlike commuting linear operators, it is not guaranteed that an eigenvector of $H_{\mathcal{A}}$ will also be an eigenvector of $\mathcal{A}$. We know that for those eigenvectors that are shared, the \emph{corresponding} eigenvalue is real. A simple $3\times 3$ example that shows how this works
\begin{equation*}
H_{\mathcal{A}}\left(\begin{array}{c}a\\b\\c\end{array}\right)
 = \left(\begin{array}{ccc} 1 & 0 & 0\\ 0 & i & 0\\ 0& 0 & -i\end{array}\right)
 \left(\begin{array}{c}a\\b\\c\end{array}\right), \quad\quad
\mathcal{A}\left(\begin{array}{c}a\\b\\c\end{array}\right)
 = \left(\begin{array}{ccc} 1 & 0& 0\\ 0 & 0 & 1\\0 & 1 & 0\end{array}\right)
 \left(\begin{array}{c}a^*\\b^*\\c^*\end{array}\right).
 \end{equation*}
 While the matrices shown do not commute, the operators \emph{do}. We see that
 \begin{equation}
 H_{\mathcal{A}} \mathcal{A} \left(\begin{array}{c} a\\b\\ c\end{array}\right) =  \left(\begin{array}{c} a^*\\ic^*\\ -ib^*\end{array}\right) =  \mathcal{A}H_{\mathcal{A}}\left(\begin{array}{c} a\\b\\ c\end{array}\right) \Rightarrow [H_{\mathcal{A}}, \mathcal{A}] = 0.
 \end{equation}
It is trivial to see that $\mathcal{A}$ is non-singular and hence has no zero eigenvalue.  We see that $\mathcal{A}$ and $H_{\mathcal{A}}$ only share one eigenvector, corresponding to the only real eigenvalue of $H_{\mathcal{A}}$. 

\paragraph{}
We see that we are guaranteed a real spectrum for $H_{\mathcal{A}}$ if in addition to the above conditions \emph{every} eigenvector of $H_{\mathcal{A}}$ is also an eigenvector of $\mathcal{A}$. This condition is referred to as \emph{unbroken} $\mathcal{A}$-symmetry \cite{introPT, discussPT, Bender:2003ve}. I make no claim that this is a necessary condition, only that it is sufficient.
\paragraph{}
Just because the Hamiltonian has real eigenvalues does not imply that the system gives rise to a sensible physical interpretation. The theory is also required to have an inner product so that amplitudes may be defined, and then probabilities derived from these amplitudes. The use of amplitudes, probabilities as the modulus of the amplitude squared and the principle of superposition are unchanged in \ptqm. The big change is that the inner product must be dynamically determined \cite{introPT, discussPT}. The way this is achieved is through introducing a basis, and defining a so-called \c\pt{} inner product by
\begin{equation}
\langle \phi_1 \vert \c \vert \phi_2 \rangle = \sum_{i,j} (\langle \phi_1 \vert)_i \c_{ij}(\vert \phi_2 \rangle)_j.
\end{equation}
In this way we can fix the formula and allow \c{} to be a dynamic object, and thus ``solved for''. Different \pt-symmetric systems require different matrices \c. Details on how to find \c{} for various Hamiltonians are given in \cite{discussPT}, but showing how to construct it is not critical for the results of this paper.

\paragraph{}
Up to this point the results are widely agreed upon and nothing new has been presented. From here, the results and discussion are new, or a prior discussion has not been found. It should be noted that the introduction of the \c{} operator introduces some possible confusion as we effectively have \emph{two} inner products in our theory: a physical (or \c\pt) inner product $\langle \phi_1 \vert \c \vert \phi_2 \rangle$ and an inner product defined by the naive ``dot product'' in terms of the components of the basis vectors $\langle \phi_1 \vert \phi_2 \rangle$. The existence of these two inner products allows a matrix representing a Hermitian operator (which is Hermitian independent of basis) to be represented by a non-Hermitian matrix. These issues are clarified in appendix \ref{app:hermite}.

\subsection*{Constructing \pt-symmetric Hamiltonians}
One way of implementing the condition $[\hpt, \pt] = 0$ with the conditions $\p^2 = \t^2 = 1$ is by the consistency condition
\begin{equation}
\hpt = \p\t\hpt\t\p\label{eq:hermite_consistency}
\end{equation}
Note that we can \emph{choose} to represent \t{} by complex conjugation: $\t(v) = v^*$. With this representation for \t{}, we check the above condition \eqref{eq:hermite_consistency} acting on an arbitrary vector $v$:
\begin{align}
\hpt(v) &= \pt\hpt\t(\p(v)) = \pt(\hpt\p^*v^*) = (\p \hpt^* \p) (v)
\end{align}
and the consistency condition is thus
\begin{equation}
\hpt = \p\hpt^* \p.\label{eq:pt_consistent2}
\end{equation}
Notice that this condition does \emph{not} ensure that \pt-symmetry is unbroken (i.e. it does not ensure that any eigenvector of \hpt{} is also an eigenvector of \pt{}.) Implementing this condition is non-trivial, and will not be attempted here. The other important point to note is that $\p{}$ is highly non-unique, and almost any $\p{}$ that satisfies $\p^2 = 1$ will do. 

\subsection*{Physically acceptable Hamiltonians}
The claim in the \pt-symmetric literature is that having a system that has unbroken \pt-symmetry is sufficient to have a reasonable quantum system. This is actually insufficient, as the following simple $2\times 2$ example shows. Choose to represent $\t$ by complex conjugation, and the parity operator by
\begin{equation}
\p = \left(\begin{array}{cc}
1 & 1\\
0 & -1
\end{array}\right).
\end{equation}
This is an acceptable parity operator as $\p^2 = 1$, although it is not diagonalisable. The Hamiltonian
\begin{equation}
\hpt = \left(\begin{array}{cc}
1 & 5i\\
0 & 1
\end{array}\right)
\end{equation}
satisfies the condition \eqref{eq:pt_consistent2}. Some key points to note:
\begin{itemize}
\item \pt-symmetry is unbroken here; ${{0}\choose{1}}$ is the only linearly independent eigenvector of either \pt{} or \hpt.
\item As a consequence, the only eigenvalue of \hpt{} is real. 
\item The eigenvectors do not span the space.
\end{itemize}
The final point is what eliminates this potential Hamiltonian as a reasonable Hamiltonian for a physical system. It is possible to find states outside the subspace spanned by the eigenvectors, and thus there is some finite probability of not getting any result for the energy at all! Another way of stating this is that this Hamiltonian, while being an example of unbroken \pt-symmetry, would not have the sum of probabilities preserved under time-evolution and so is non-unitary.

\paragraph{}
In the literature, it is always assumed that the matrix $\p$ can be chosen to be diagonal without loss of generality. In that case the eigenvectors of \pt{} span the space, and the unbroken requirement ensures that the eigenvectors of the Hamiltonian also span the space. But as it was shown that the $\p$ cannot always be diagonalised (as above), I will replace the condition ``unbroken'' with what I believe to be conditions for any physically reasonable Hamiltonian:
\begin{enumerate}
\item \emph{Real eigenvalues:}\\ These are the results of measurements, and so they have to be real.
\item \emph{Diagonalisable:}\\ If the eigenvectors of $H$ span a \emph{finite} Hilbert space, then $H$ will be diagonalisable.
\item \emph{Eigenvectors are orthogonal:}\\ This is a requirement on the inner product. If our system is in an eigenstate of an observable, then a measurement of this system is guaranteed to return that value for the observable. Therefore the probability of getting any other result must be zero. Therefore the inner product between any two eigenvectors of the same observable must be zero.\footnote{If the observable is degenerate, then one can perform a Gram-Schmidt procedure on the degenerate eigenspace.}
\item \emph{Probabilities must add to one}\\
The evolution must be ``unitary'', in the sense that the probabilities add to one. Note this is a condition on both the inner product and the Hamiltonian.
\end{enumerate}
These conditions define a physically reasonable version of quantum mechanics, regardless of whether the Hamiltonian is Hermitian, \pt{}-symmetric or something else.

\paragraph{}
Of particular interest in this paper are two dimensional systems, as they are simple to analyse explicitly. If \p{} is chosen to be the matrix
\begin{equation}
\p = \left(\begin{array}{cc} 0 & 1\\ 1 & 0\end{array}\right)
\end{equation}
then the most general solution to \eqref{eq:pt_consistent2} is
\begin{equation}
\hpt = \left( \begin{array}{cc} h_{11}  & h_{12}\\
h_{12}^* & h_{22}^*\end{array}\right)\label{eq:2x2form}
\end{equation}
where $h_{ij}$ are arbitrary complex numbers. It is important to note that just because a matrix is not of this form does \emph{not} imply that the system does not possess $\pt{}$ symmetry. This is the most general $2\times 2$ \pt-symmetric Hamiltonian \emph{for this choice of }\p. 

\section{Hermitian problems to \ptqm}
\label{sec:QMNO}
In this section we will show explicitly how to construct quantum mechanics of a spin-1/2 particle in a magnetic field in a non-orthogonal basis and show that this leads to a non-Hermitian Hamiltonian with a modified inner product. Taking the magnetic field to act in the $x$-direction, we have
\begin{equation}
H = \varepsilon(\up \downC + \down \upC) = \varepsilon \left(\begin{array}{cc} 0 & 1\\ 1 & 0\end{array}\right)
\end{equation}
when expressed in the conventional (orthogonal) basis $\up= {{1}\choose{0}}$, $\down = {{0}\choose{1}}$.

\paragraph{}
While it is unorthodox, there is no reason why this system cannot be described in a non-orthogonal basis. As a basis transformation is a mathematical transformation only, it cannot affect any observable property of the system but it will change the description of the system. As a trivial example, the \emph{same} vector will generically have different components in a different basis. Another example is that preserving the inner product between two vectors will require the \emph{formula} for the inner product in terms of the components must also change. Denoting quantities in the non-orthogonal basis with a prime we can implement a change of basis in the following way:
\begin{equation}
\vert \phi^\prime \rangle = B^{-1} \vert\phi\rangle.
\end{equation}
To obtain a Hamiltonian $H^\prime$ of the form \eqref{eq:2x2form} a basis transformation with a free parameter $\alpha$ is chosen
\begin{equation}
B^{-1} = \left(\begin{array}{cc}
\cos\alpha   &-i\sin\alpha\\
-i\sin\alpha & -\cos\alpha\end{array}\right)\label{eq:basis_choice}.
\end{equation}
As this is not a transformation between orthonormal bases, $B^{-1}$ is not unitary or anti-unitary. This transformation is well defined provided that $\alpha \neq \frac{\pi}{4} + 2\pi n$, as at these values of $\alpha$ the basis vectors are linearly dependent. 

\paragraph{}
Because this is only a change in basis, the Hamiltonian must transform in the following way:
\begin{equation}
H^\prime = B^{-1} H B = \frac{\varepsilon}{\cos 2\alpha}\left(\begin{array}{cc}
-i\sin 2\alpha & -1\\
-1 & i\sin 2\alpha
\end{array}\right).
\end{equation}
Because this a basis change, the eigenvalues must remain unchanged, as can be explicitly checked. The matrix $H^\prime$ is not Hermitian (in the sense that it is not equal to the transpose of its complex conjugate). It is \pt{}-symmetric, in the sense that it is of the form \eqref{eq:2x2form}.

\paragraph{}
Note that this procedure is well defined for any non-singular matrix $B$ and must be a valid description of the system. However not every matrix $B$ will end up with $H^\prime$ being of the form \eqref{eq:2x2form}. As discussed immediately after \eqref{eq:2x2form} this does not necessarily mean that $H^\prime$ is not \pt-symmetric, but may correspond to a different choice for \p. The other reason for making the choice \eqref{eq:basis_choice} is that by adjusting $\alpha$ we are able to show how the quantum brachistochrone problem is resolved.


\subsubsection*{Inner product}
As we are only changing our description of the system by changing basis, we must preserve the inner product between vectors. As noted above, because the same vectors have different components in the new basis it is necessary to change the formula for the inner product in terms of the components. Explicitly, we require
\begin{equation}
\langle \psi \vert \phi \rangle = (B\vert\psi^\prime\rangle)^\dagger(B\vert\phi^\prime\rangle) 
= \langle \psi^\prime \vert B^\dagger B \vert \phi^\prime \rangle
\end{equation}
If we were transforming between orthogonal bases then $B$ would be (anti-)unitary, so $B^\dagger B = (-)1$. This would preserve the inner product. For a non-orthogonal transformation, we see the role of $B^\dagger B$ is the same as $\c$ in \ptqm. For the case under consideration
\begin{equation}
\c = B^\dagger B = \frac{1}{\cos^2 2\alpha}\left(\begin{array}{cc}
1 & -i\sin 2\alpha\\
i\sin 2\alpha & 1\end{array}\right).
\end{equation}

\subsubsection*{Time evolution}
The system is known to have unitary time evolution in the original basis, and the time evolution is completely specified. In the Schr\"odinger picture we have
\begin{equation}
\vert\phi(t) \rangle = U(t,0) \vert \phi(0) \rangle, \quad\quad U(t,0) \equiv \exp(-iHt/\hbar).
\end{equation}
As only our description of the system has changed, the probabilities still must add to one when described in the new basis. The time evolution operator in the new basis takes the form
\begin{align}
U(t,0)^\prime &= B^{-1} U(t,0) B = B^{-1} \left(\sum_j \frac{(-it)^j}{\hbar^j}(HBB^{-1})^j\right)B = \exp(iH^\prime t/\hbar)\nonumber
\end{align}
Thus $U(t,0)^\prime$ takes the form we would expect for a Hamiltonian. Because the matrix $H^\prime$ is not Hermitian, the matrix $U(t,0)^\prime$ is also not unitary. As explained above the time evolution operator \emph{is} unitary, it is just not represented by a unitary matrix because the matrix elements are defined with respect to a non-orthogonal basis. This point is addressed further in appendix \ref{app:hermite}.

\subsubsection*{Beating the brachistochrone}
Recently there has been a suggestion that \ptqm{} allows for faster evolution than Hermitian quantum mechanics, and that this could form an experimental difference between \pt-symmetric and ordinary quantum mechanics \cite{FastBender}. More precisely, let us start with two states ${{1}\choose{0}}$ and ${{0}\choose{1}}$. We then choose a Hamiltonian which minimises the time taken to evolve from ${{1}\choose{0}}$ to ${{0}\choose{1}}$ subject to the constraint that the difference between the highest and the lowest eigenvalue of $H$, denoted $\omega$, remains fixed. This is known as the quantum brachistochrone problem. It was shown that the fastest a \emph{Hermitian} Hamiltonian can evolve from  ${{1}\choose{0}}$ to ${{0}\choose{1}}$ is $\tau = \pi \hbar/\omega$ \cite{quantumshort}. In contrast, $\tau > 0$ is the only constraint for a \pt-symmetric Hamiltonian \cite{FastBender}. If we experimentally discover a situation where a system is evolving faster than allowed by Hermitian quantum mechanics, then the claim is that it may still be described by \ptqm. This is in direct conflict with the claim of this paper which is that for finite dimensional systems \ptqm{} is ordinary quantum mechanics in a non-orthogonal basis, and therefore equivalent. 

\paragraph{}
It is instructive to work through the spin-1/2 example and show that there is no experimental test here. To do this, let us calculate how much time it takes to evolve from $\vert e_1^\prime \rangle = {{1}\choose{0}}$ to $\vert e_2^\prime \rangle = e^{i\lambda}{{0}\choose{1}}$ in the non-orthogonal basis, where we are not concerned about the overall phase.\footnote{If you wanted to answer the original brachistochrone problem, then you would need to have defined additional phase factors in the definition of $B$. In order to keep the presentation as intuitive as possible this was not done.} We can carry out this calculation in two ways: work out what the vectors $|e_1 \rangle$ and $|e_2\rangle$ are in the original basis and use normal quantum mechanics to evolve the system, or use the time evolution operator $U(t,t_0)^\prime$. The time taken is given by
\begin{equation}
\tau = \frac{\hbar}{\varepsilon}\tan^{-1}\left(\frac{1}{\tan 2\alpha}\right)
\end{equation}
We see as $\alpha \rightarrow 0$ that the minimum time approaches the brachistochrone limit, $\tau \rightarrow \hbar\pi/2\varepsilon$. This is no surprise, as in this limit $H^\prime$ is Hermitian. But as $\alpha \rightarrow \pi/4$ we see $\tau \rightarrow 0$!

\paragraph{}
To get some understanding for what is happening, it is useful to transform back into the original basis. Doing this tells us
\begin{align}
|e_1\rangle &= B|e_1^\prime\rangle = \frac{1}{\cos2\alpha}\left(\begin{array}{c} -\cos\alpha \\ i\sin\alpha\end{array}\right)\\
|e_2\rangle &= B|e_2^\prime\rangle = \frac{1}{\cos2\alpha}\left(\begin{array}{c} i\sin\alpha \\ \cos\alpha\end{array}\right)
\end{align}
Once written this way it becomes apparent that $\alpha \rightarrow \pi/4$ corresponds to taking the two basis vectors to be almost degenerate. As long as the basis vectors are not collinear, they may become as close as desired, and so the minimum time to evolve from one to the other is bounded by zero. The notion of \ptqm{} being faster than Hermitian quantum mechanics comes about from comparing the amount of time taken to evolve between states with given \emph{components}, rather than between given states. 

\section{General finite Hamiltonians}
\label{sec:PT2QM}
We have seen that we can transform ordinary quantum mechanics problems into \pt{}-symmetric quantum mechanics problems by choosing a non-orthogonal basis. The question remains if there is any \emph{physically acceptable} \pt{}-symmetric Hamiltonian that is not ordinary (i.e. not just Hermitian Hamiltonians in a different basis). If not, then the \pt{}-symmetric quantum mechanics is simply a change in description of known physics. 

\paragraph{}
The proof that the physically acceptable \emph{finite} \pt-symmetric Hamiltonians are equivalent to ordinary quantum mechanics is surprisingly short:
\begin{enumerate}
\item The eigenvectors must span the Hilbert space, and this is a sufficient condition that the Hamiltonian can be diagonalised by a similarity transformation.
\item The eigenvalues are real, so in diagonal form the Hamiltonian is obviously Hermitian.
\item In the basis in which the Hamiltonian is diagonal, the eigenvectors are of the form
\begin{equation*}
(\vert E_i \rangle)_j = \delta_{ij}
\end{equation*}
Because we require that the eigenvectors are orthogonal, this fixes the inner product to be the standard one in the diagonal basis.\footnote{Actually, it is only important that eigenvectors that correspond to \emph{distinct} eigenvalues are orthogonal. For the degenerate case, one can perform a Gram-Schmidt procedure once an inner product is defined.}
\end{enumerate}
But at this point we have a Hermitian Hamiltonian and a standard inner product -- we have recovered ordinary quantum mechanics! Therefore the standard proof that time evolution is unitary now follows.

\paragraph{}
In various papers the claim is made that one of the properties that makes \ptqm{} novel is that the physical  inner product  (i.e \c) is determined dynamically. From the viewpoint of physically reasonable systems, the explanation for this is straightforward. By choosing to implement a Hermitian Hamiltonian the conditions are automatically satisfied. By allowing more freedom in the form of the original Hamiltonian the constraints are not automatically satisfied, and the extra freedom is compensated for by imposing constraints on \c.

\section{Further research}
\label{sec:further}

\subsubsection*{\pt-symmetric field theory}
It has been shown how changing bases can transform a Hermitian quantum system into a non-Hermitian one. This can be done regardless of whether the system in question is finite, countably infinite or even a field theory and therefore some \pt{}-symmetric field theories are equivalent to ordinary quantum mechanics. The results presented in the reverse direction are applicable only for finite systems, although some work has been done on the general case. In \cite{basisProof}, it was shown that there existed a matrix $B$ such that 
\begin{equation}
H_{\mathcal{PT}} = B H_{\textrm{Hermitian}} B^{-1}
\end{equation}
The proof given in \cite{basisProof} also shows how to construct such a matrix $B$ (although note that $B$ is highly non-unique, as $B\rightarrow BU$ for any (anti-)unitary matrix $U$ will also suffice). While the proof is not complete without also looking at the inner product, these first results seem to indicate that \pt-symmetric quantum field theory is ordinary field theory in a different basis.

\paragraph{}
If \pt-symmetry does not introduce any new physics, it should be asked why should we study it at all? The answer is one of pragmatism: sometimes working in a different basis makes the problem simpler and more tractable. A  good example is given in \cite{simpleCubic}, where the \pt-symmetric Hamiltonian density
\begin{equation}
\mathcal{H}_{\pt} = \frac{1}{2}p^2 + \frac{1}{2}x^2 + igx^3
\end{equation}
is perturbatively compared to its Hermitian counterpart, which involves more interaction terms and hence more Feynman rules. It is also shown in this paper that the Hermitian version possesses divergences absent in the equivalent \pt-symmetric theory. For this theory the \pt-symmetric version is much easier to work with.

\paragraph{}
It may seem odd to claim that theories are equivalent, yet have divergences in the Hermitian form of the theory while such divergences are absent in the \pt-symmetric formulation. However this is not a contradiction! There are two separate issues: the background being perturbed about and the ambiguity in the quantisation process. Taking a well defined quantum theory and quantising around an unstable solution leads to a perturbative theory which appears to have energies unbounded below. Such a problem would be apparent quantising the Higgs potential $\lambda (|\phi|^2 - v)^2$ around $\phi = 0$. The other issue that occurs is that the quantisation procedure itself is ambiguous. It seems that the quantisation procedure for the cubic theory discussed in \cite{simpleCubic} seems to pick out a good quantisation procedure and background whereas a \emph{naive} quantisation procedure in the Hermitian version leads to divergences. It would be an interesting research project to quantise the equivalent Hermitian Hamiltonian in a way equivalent to the \pt-symmetric version of the theory. 

\subsubsection*{Develop Hamiltonian intuition}
Many Hamiltonians of interest in field theory come about starting with a classical system and then quantising it. In choosing potentials to model a particular situation we draw on our intuition of how the corresponding classical potential would cause the system to behave. Part of the interest in \ptqm{} is that the Hamiltonians look like they would have ill-defined classical limits, such as the Hamiltonian $\hpt = p^2 - x^4$. While \pt-symmetry may make the theory simple to solve, it obscures the classical limit. 

\paragraph{}
We can take an explicit example from the work of Bender et al \cite{simpleCubic}. The \pt-symmetric Hamiltonian is given by
\begin{equation}
\hpt = \frac{1}{2} p^2 + \frac{1}{2}x^2 + ix.
\end{equation}
This Hamiltonian is equivalent to the Hermitian Hamiltonian \mbox{$H = \frac{1}{2}(p^2 + x^2 + 1)$}. As they are equivalent theories, they both have the same physical interpretation. A simple way of obtaining the classical limit of a \pt-symmetric theory would aid the physical interpretation of the theory. 

\subsubsection*{Develop a method for choosing physically reasonable Hamiltonians}
Finding if a theory is invariant under simultaneous spatial inversion and time reversal is fairly simple. The condition that it is physically reasonable is not. It is generally hard to show that all the eigenvectors of $H$ are also eigenvectors of $\pt$, thus ensuring their reality. This has been done in specific cases, but I know of no general procedure to show if a given Hamiltonian is reasonable. 

\section{Conclusions}
This paper has clarified the role of \pt-symmetric quantum mechanics, and shown that it comes about from a rewriting of quantum mechanics in a non-orthogonal basis. As a consequence, there is no experimental test that could ever distinguish between Hermitian quantum mechanics and \ptqm. It is shown how the recent claim that \ptqm{} allows for faster evolution is a coordinate artefact.

\paragraph{}
Because the ideas presented here are relatively simple, it is worth asking to what extent they are already known. The final paragraph of \cite{simpleCubic} has an interesting discussion that would suggest that this is already known. In particular, Mostafazadeh has argued that ``A consistent \pt-symmetric quantum theory is doomed to reduce to ordinary quantum mechanics'' \cite{mos2}. Bender et al acknowledge that ``Mostafazadeh appears to be correct'', but argue that as a pragmatic issue that in some cases a \pt-symmetric Hamiltonian leads to a much simpler quantisation scheme. There is no argument with this result. But to add to the confusion, one of the authors then proposed an \emph{experimental} test to distinguish between \ptqm{} and ordinary quantum mechanics. This paper seems worthwhile as it appears that there is still confusion in the field about the status of \ptqm.

\paragraph{}
This paper also serves a pedagogical purpose, namely it provides an elementary proof for finite dimensional systems that observables must be represented by Hermitian operators. Appendix \ref{app:hermite} shows that a Hermitian operator does not have to be represented by a Hermitian matrix, provided that the inner product is modified. Currently most undergraduate quantum mechanics books prove that Hermitian matrices have the desired properties enumerated for physically reasonable theories, but the results presented here demonstrate that there is no loss of generality in taking a physically reasonable observable to be Hermitian.

\subsection*{Acknowledgements}
I would like to thank Nemanja Kaloper and Joe Kiskis for illuminating discussions and suggesting extensions to this work. I would also like to thank Jamison Galloway for careful proof-reading.


\appendix

\section{Hermitian operators and basis changes}
\label{app:hermite}
It is claimed that by changing basis we can transform a Hermitian matrix into a non-Hermitian one, and show how \ptqm{} arises from writing quantum mechanics in a non-orthogonal basis. An operator $H$ on a Hilbert space is Hermitian if it is everywhere defined and satisfies $\langle H \phi_1 \vert \phi_2 \rangle = \langle  \phi_1 \vert H\phi_2 \rangle$ for all $\vert \phi_i \rangle$. However, whether or not an operator is Hermitian does depend on the inner product defined on the space. 

\paragraph{}
Because the operators representing observables are Hermitian in the original basis, the operators must be Hermitian. The reason that these Hermitian operators can be represented by non-Hermitian matrices is that we have two inner products:
\begin{enumerate}
\item An inner product defined directly in terms of the components of the basis vectors, $\langle  \phi_1 \vert \phi_2 \rangle$. Because of the way it is defined, this inner product explicitly depends on the basis chosen.
\item A physical (or \c\pt) inner product used to compute amplitudes $\langle \phi_1 \vert \c \phi_2\rangle$. 
\end{enumerate}
It is important to note that there is no assumption that the Hilbert space is endowed with an inner product. Instead the first inner product is fixed for convenience, and a general inner product is allowed for by introducing $\c$. Then ``it [the inner product, or equivalently \c] is dynamically determined by $H$'' \cite{scalarQFT}. As explained at the end of section \ref{sec:PT2QM} it is more accurate to think of \c{} being constrained so the resulting theory has a reasonable interpretation.

\paragraph{}
The resolution of how a Hermitian operator is represented by a non-Hermitian matrix is that the matrix elements of a matrix in a particular basis is defined by the first of these inner products. In this (unphysical) inner product physical observables are generically not Hermitian. Thus when an operator is written in matrix form, it does not appear as a Hermitian matrix. In the second (physical) inner product, the observables \emph{are} still Hermitian. If this inner product were used to define the matrix elements, the resulting matrices would be Hermitian, thus removing the apparent contradiction originally posed.

\paragraph{}
Similar remarks hold for the term unitary. The time evolution operator $U(t,t_0)^\prime$ introduced in the paper is not unitary with respect to the first inner product (and hence in terms of its matrix representation) if quantum mechanics is done in a non-orthogonal basis. However this operator is still unitary in the physical inner product. Because of this slight semantic confusion, the fourth condition for reasonable time evolution is stated as probabilities must always add to one, rather than the more common phrase ``time evolution is unitary''.


\end{document}